\documentstyle[12pt,epsfig]{article}
\begin{document}
\begin{titlepage}
\begin{center}
\vspace*{2cm}

{ \Large  \bf  ON THE HIGH ORDER \\~
\\ MULTIPLICITY MOMENTS } \vspace{2cm}

\begin{author}
\Large K. Fia{\l}kowski\footnote{e-mail address:
uffialko@th.if.uj.edu.pl}, R. Wit\footnote{e-mail address:
wit@th.if.uj.edu.pl}

\end{author}

\vspace{1cm}

{\sl M. Smoluchowski Institute of Physics\\ Jagellonian University \\

30-059 Krak{\'o}w, ul.Reymonta 4, Poland}

\vspace{2cm}

\begin{abstract}
The description of multiplicity distributions in terms of the
ratios of cumulants to factorial moments is analyzed both for data
and for the Monte Carlo generated events.  For the
PYTHIA generated events the moments are investigated for the
restricted range of phase-space and for the jets reconstructed from
single particle momenta. The results cast doubts on the validity of
extended local parton-hadron duality and suggest the possibility of
more effective experimental investigations concerning the origin of
the
 observed structure in the dependence of moments on their order.

\end{abstract}

\end{center}
\vspace{1cm}



\vspace{0.5cm}

\end{titlepage}

\section{Introduction}
   ~~~~For decades the investigation of the moments of multiplicity distributions for high energy
    hadroproduction processes has been limited to the lowest three orders. It was generally believed
    that the experimental errors make the results for the fourth and higher cumulants meaningless. The
    investigations of higher normalized factorial moments in terms of intermittency \cite{BP} was possible by
    means of averaging the distributions over many bins in phase space, thus efficiently increasing
    the statistics. Still, even in this case the fourth and higher moments seemed to be
    determined to large extent by the values of the first three moments.
\par   The breakthrough came with the analysis of Pavia group \cite{Pavia}, who have shown that the analysis can be
   significantlly extended for a new kind of moments: the ratios of cumulants to the factorial moments
   $$H_q = K_q/F_q.$$
    Here the factorial moments are defined in the standard way
    $$F_q=\sum_{n} \frac {n!}{(n-q)!}P(n),$$
and the cumulants may be calculated from a recursive formula
$$K_q=F_q-\sum_{i} \frac{(q-1)!}{(i-1)!(q-i)!}K_{q-i}F_i.$$
For
 the theoretical models it is sometimes easier to calculate moments
from the generating function of the multiplicity distribution
$$g(z) = \sum_{n} z^n P(n).$$
 The corresponding formulae read
 $$F_q=\frac{d^q}{dz^q}g(z)|_{z=1};~~~K_q=\frac{d^q}{dz^q}\log g(z)|_{z=1}.$$
 If the values of the $F_q$ moments of the order $q>3$ are determined by the values of lower order moments,
 the values of the $K_q$ moments are consistent with zero. In the Pavia group analysis the $H_q$ moments
 were shown to be significantly different from zero for $q=4,~5$ and $6$, and possibly also for $q>8$.
The reason why
 the new moments of higher orders seem to have much smaller
relative errors than the cumulants of the same order is
the cancellation of some contributions to the overall errors.
Fluctuations in the high multiplicity tail affect in a similar way
the cumulants and factorial moments, and the resulting
fluctuations of the values of their ratios are damped.
\par
   The behaviour  of the moments found by the Pavia group initiated much interest because of their apparent 
   compatibility with the
    predictions of perturbative QCD at the NLLA level \cite{D1},\cite{D2}. In both cases the dependence
    of moments on their order
    was found  to be non-monotonical: after a minimum (with negative value) at $q=5$, a hint of oscillations
    at higher $q$ was seen. Later, however, some doubts appeared about the
    origin of the observed structure. In particular, a simple cut in the smooth multiplicity
    distribution (e.g. of the negative binomial type) was shown to produce similar effects \cite{UGL}.
   In this note we analyze the behaviour of the moments calculated from the PYTHIA generator \cite{PYTHIA}
   for
 the electron-positron annihilation at LEP-I energy and compare them, where possible, with the
   experimental L3 results \cite{L3}, \cite{MA}. We include also some proposals for future investigations.

\section{Data and Monte Carlo results for Z decays}
~~~~ To measure reliably higher multiplicity moments, one needs
high statistics and negligible systematic
     errors. The LEP-I data, where millions of events have been collected by each experiment using detectors covering
     almost $4\pi$ solid angle with very high efficiency, seem to be ideal for this purpose. The results of L3
     collaboration \cite{L3} were published recently (and compared with JETSET \cite{J} and HERWIG \cite{H}
      Monte Carlo results).
     In this section we remind these results and compare them with the moments calculated from PYTHIA.
     The comparison is only qualitative, as we do not use the L3 detector simulation program, which served
     to unfold the experimental multiplicity distribution \cite{L3}. Nevertheless, we will see that some
     surprising features of data and MC results are confirmed.
 We should note also that the overall event
     characteristics is properly described by PYTHIA: e.g., the values of the average charged multiplicity 
     (above $18$) and of the dispersion (about $5.9$) are reproduced correctly. 
     
   \par
   The values of $H_q$ moments calculated from the unfolded multiplicity distributions are reproduced in
   Fig.1a for the order $q=4\div 19$ together with the expectations of JETSET (the points for $q=2,3$
   are never shown in such plots, because they lie much higher). It is obvious that
   JETSET results agree with data well (we do not reproduce here the unsuccessful comparison of data with
   HERWIG).
    However, it is equally obvious that the errors
 read out directly from the figures 
   seem overestimated. The JETSET results (obtained without tuning the parameter values) never deviate
   from data by more than one standard deviation, and for the majority of
   moments agree with data within less than half of SD. This is statistically improbable and suggests that the
   values shown cannot be interpreted as simple uncorrelated statistical errors. Indeed, they are just the
   diagonal elements of the covariance matrix which is influenced by large bin-by-bin correlations.
   \par
    The authors seem to recognize this problem, and they do not use
   directly the multiplicity distributions obtained from the data by an unfolding procedure.
    Instead they use the multiplicity distributions truncated
   in such a way, that multiplicities with relative
   error on $P(n)$ greater than $50 \% $ are rejected.   This corresponds to the cut at highest
   multiplicities removing about $0.035 \% $ of events. With such a truncation the errors are
   visibly reduced and the clear structure appears both in data and MC results shown in Fig.1b:
   a minimum with negative values of $H_q$ at $q=5\div6$, and possible oscillations with maxima for $q$ around
   $10\div11$ and $16\div17$.
   
        \begin{figure}[h]
\centerline{\epsfig{figure=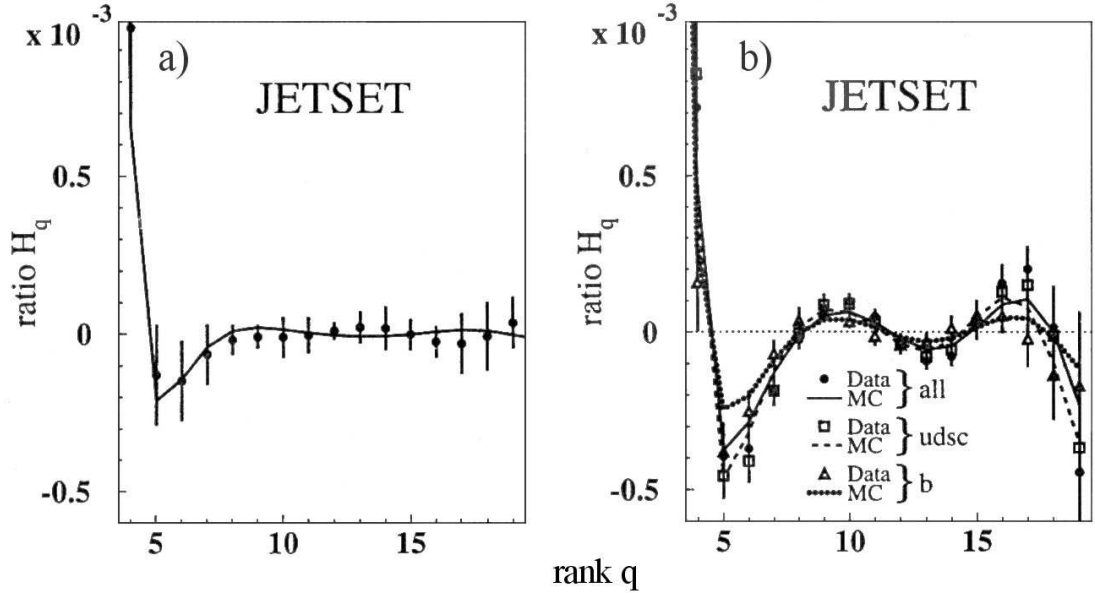, height =8cm}}
\caption{\label{fig1} {\small \sl The L3 results \cite{L3} for $H_q$ moments calculated from full (a) and truncated
 (b) multiplicity distribution. Black points represent the data, and solid curve the events generated from JETSET.
 In Fig. 1b additional points and curves correspond to the separated $udsc$ and $b$ quark contributions.}}
\end{figure}

    \par
     In our opinion the lack of fluctuations above one standard deviation and the reduction of errors after the
     removal of some data suggest strongly that the error estimate  may be misleading. Thus for our simulations
     using PYTHIA we do not calculate errors from standard statistical formulae (relating the errors of the
q-th moment to some moments of order $2q$), but use instead the
uncertainty evaluated from the spread of results for different
samples of independently generated events.
 We use the statistics, for which the errors are comparable to the
size of symbols in the figures.
   \par
   The results for samples of $4$ million  events (more than twice the experimental sample of L3)
   are shown in Fig. 2. We have checked that the choice of PYTHIA parameters (default values,
   or the values tuned to L3 data) does not influence significantly our results. 
   We restricted our calculations to $q<11$, as the results for
   higher $q$ seem unstable even for such a huge statistics. We confirm the shallow minimum in the untruncated data,
   which becomes deeper after introducing the cut. Cutting off bigger part of the high multiplicity tail makes
   the minimum still more pronounced. 
   
 \begin{figure}[h]
\centerline{\epsfig{figure=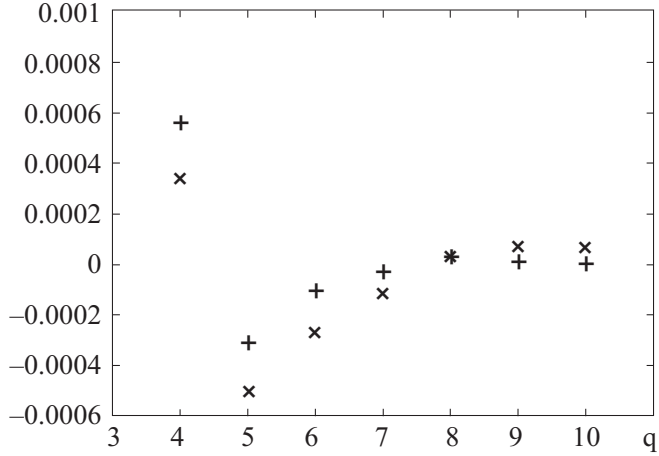, height =6cm}}
\caption{\label{fig2} {\small \sl The $H_q$ moments calculated
from PYTHIA events for  L3 parameters
. Crosses represent full distributions and $x$-s the
distributions with the highest multiplicity tail (0.035\% of cross
- section) removed.}}
\end{figure}
   
   \par
   We checked that the minimum in untruncated data does not change significantly if we change the number of
   generated events between $2$ and $4$ million; thus it is not due to the "natural cut" resulting form the finite
   statistics. This confirms once more the results from many other investigations:
   the existence of minimum
 seems to be universal, but its shape depends strongly on the multiplicity cuts.  If the statistics  is low,
   such cuts are "naturally" present in data; for high statistics experiments they may result from detector
   deficiencies.

\par
      The origin of the presence of a structure in the dependence of the $H_q$ moments on their order $q$
 for the PYTHIA events is not clear. We think that these effects are not due to the NNLO order QCD
 perturbative effects \cite{D1}, \cite{D2}, as the PYTHIA generator does not contain explicitly such components.
 A more plausible explanation is to relate the structure more generally to the multicomponent character
 of the production process \cite{D3}. An example of such  "multicomponent expansion" for Z decays may be a separation
 of two-, three- and multijet classes of events. It seems natural to assume a smooth (e.g. negative
 binomial - NBD) distribution to the two-jet events \cite{GU}. For such a distribution characterized by two parameters,
 $\overline n$ and $k$
$$ P_n^{NBD}(\overline{n},k) =
\frac{\Gamma(n+k)}{n!\Gamma(k)}\bigg(\frac{\overline
n}{\overline{n}+k}\bigg)^n\bigg(\frac{k}{\overline{n} +
k}\bigg)^k$$
 one can easily calculate the factorial moments
$$F_q^{NBD}(\overline{n}, k) =
\bigg(\frac{\overline{n}}{k}\bigg)^q\frac{\Gamma(k+q)}{\Gamma(k)}.
$$
\par
The distributions for three- and multijet events should be also
related to NBD. In a toy model, in which only two- and
three-jet events are taken into account, and the parameters in
this two classes are scaled in the $2:3$ ratio
 we have
$$ P_n = \alpha P_n^{NBD}(\overline n, k)
+(1-\alpha)P_n^{NBD}(3\overline n /2,3k/2).
$$
Obviously, in such a toy model we get for the factorial moments 
$$ F_q = \alpha F_q^{NBD}(\overline n, k)
+(1-\alpha)F_q^{NBD}(3\overline n /2,3k/2)
$$
and from the formulae given in the Introduction one may easily
write down the explicit formulae for the $H_q$ moments of any
order.
   We have calculated these moments for $q<16$ for the wide range of model parameters.
   We found that for the values of $\overline n$ and $k$ around $20$ (as suggested by the experimental
   values of the average multiplicity and dispersion), and for $a\sim  0.85$ (corresponding to a $15\%$
   admixture of three-jet events, as suggested by data) a minimum for $q=5$ (with a value around $-0.0005$)
   appears naturally. We checked also that a cut removing highest multiplicity events (at the level of $0.01\%$)
   enhances such a minimum. Thus any generator reproducing correctly the two- and three-jet components may be expected
   to reproduce also qualitatively the observed structure in the dependence of $H_q$ moments on their order $q$.

   \par
   We conclude that the minimum in the dependence of moments $H_q$ on their order $q$ seems to occur naturally at
   $q=5\div6$ for the hadron multiplicity distributions in the full phase-space. Its shape depends, however, on the
   details of the generating procedure. In particular, the cuts removing high multiplicity tail enhance the
   minimum quite strongly.

\section{Higher moments in restricted parts of phase-space}
\par

    The data and simulations discussed above concerned the full phase-space multiplicity distributions,
    unless limited by the experimental conditions. It is well known, however, that the multiplicity
    distributions in limited regions of phase-space depend significantly on the definition of the limits.
    For the early collider data \cite{UA5} the successful negative binomial (NB) fits were found both for full
    available phase-space and for the intervals of CM rapidity. However, the $1/k$ parameter (equal to second
    normalized cumulant for NB distributions) increases significantly for a decreasing size of the rapidity interval.
    It was approximately described by the hypothesis of $\overline{n}/k$ scaling \cite{VH} or by the minimal 
    model \cite{KF},
 in
 which the only significant correlations were those reflected by the multiplicity distributions for
    full phase-space. The data were not accurate enough to measure higher order cumulants and the
    statistics was insufficient to investigate small intervals (which was done later for intermittency
    effects using the averaging over many intervals \cite{BP}).

    \par
   The LEP1 data have sufficient accuracy and statistics to investigate higher moments for various regions
   of phase-space. Thus we have performed calculations of such moments for the events generated with PYTHIA,
   defining phase-space regions by cuts in values of CM momenta. It should be interesting to see if the
   patterns revealed by such calculations will be confirmed by future data analysis.
   \par
   We have generated samples of $4M$ events both for the default values of PYTHIA parameters and for the values
   used by L3 collaboration. We calculated the moments for orders $2\div 10$ for regions defined by simple
   inequality for the values of the CM three-momenta $p$: $p<\epsilon n$, where $\epsilon =0.2GeV$, and $n=1\div 10$.
  \par
   Although the results for two choices of parameters are different, there is a common
   pattern in them; thus we show only the results for $L3$ parameters.
    The minimum (with negative values) in the dependence of moments $H_q$ on their order
   $q$ occurs for small phase space regions very late (at $q$ around $8$) and shifts gradually to smaller
   values of $q$ for increasing range of CM momenta. However, even for the widest range of momenta investigated,
   the position and shape of minimum differs significantly from that for full phase-space.
    The results are shown in Fig.3a; for transparency we show only points for $n$ =4,6,8 and 10.

            \begin{figure}[h]
\centerline{\epsfig{figure=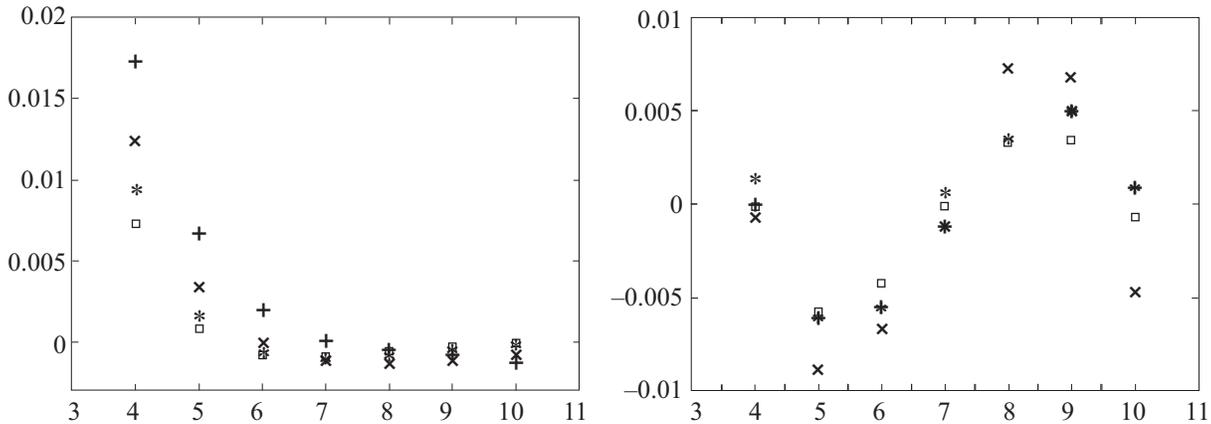, height =5.5cm}}
\caption{\label{fig3} {\small \sl The  $H_q$ moments calculated
from PYTHIA events in limited parts of phase-space (see text) for
 L3 values of the parameters without (a) and with (b) a cut removing highest multiplicities.
  Crosses, $x$-s, stars and squares represent the results for $n$ = 4, 6, 8 and 10,
respectively.}}
\end{figure}

   \par
    In the previous chapter we have seen that the minimum for full phase-space was strongly enhanced by
    introducing cuts in the multiplicity distributions, although for the LEP1 data the minimum cannot be
    explained just by such a naturally occurring cut due to the finite statistics. Thus we repeated our
    calculations cutting off the highest multiplicities contributing $0.001$ to the full cross-section.
    The results, shown in Fig. 3b, are quite surprising: now the position of minimum is practically
    independent on the size of phase-space region selected. It will be very interesting to see if the
    data show the same patterns when investigated with- and without extra cuts in multiplicity.

\section{Higher moments for hadrons and for jets}

\par
   As noted in the introduction, the existence of minimum (with negative values) in the $q$-dependence
   of moments $H_q$ was related to the higher order corrections in
   perturbative QCD. It is obvious, however, that the perturbative QCD calculations yield the
   multiplicity distributions for gluons, and not for single hadrons. The identifications of
   higher moments for those two distributions means a rather bold extension of the assumption of
   parton-hadron duality, usually applied only to the average values.
   \par
To estimate the reliability of such an extension we applied to the
generated events the default clustering algorithm of PYTHIA (PYCLUS) and investigated the
multiplicity distributions of reconstructed jets (which may be
expected to correspond to the gluon distribution more closely,
than the distribution of single hadrons).
    The average number of jets depends strongly on the values of two parameters. One of them defines the maximal
    phase-space distance between the particles added to the existing jet; the other one defines the maximal CM momentum
    for the  slowest particles in CM, which form a separate cluster/jet.
    \par
     We calculated the ten lowest moments
    for the distributions of jets defined with those parameters spanning a range of $0.02-0.1$ GeV, 
    for the L3 parameters, with a cut in the multiplicity. An additional degree
    of freedom was introduced by including in the clustering algorithm not only the charged stable hadrons
    (stable in the experimental sense, i.e. coming to the detectors; this is the condition used in all the distributions), but
    also stable neutral particles. These are mainly photons (coming from the ${\pi}^0$ decays), but also
    (anti)neutrons and long-living
    neutral kaons. This option allows to compare the distribution of hadrons and jets with the same
    average multiplicity.
   The results of our calculations are shown in Fig.4. It is obvious that the values of moments
 for
 jets are very different from those for hadrons, even for relatively small jets (when the multiplicity
   in both cases is similar); e.g., the fourth moment is now negative.

            \begin{figure}[h]
\centerline{\epsfig{figure=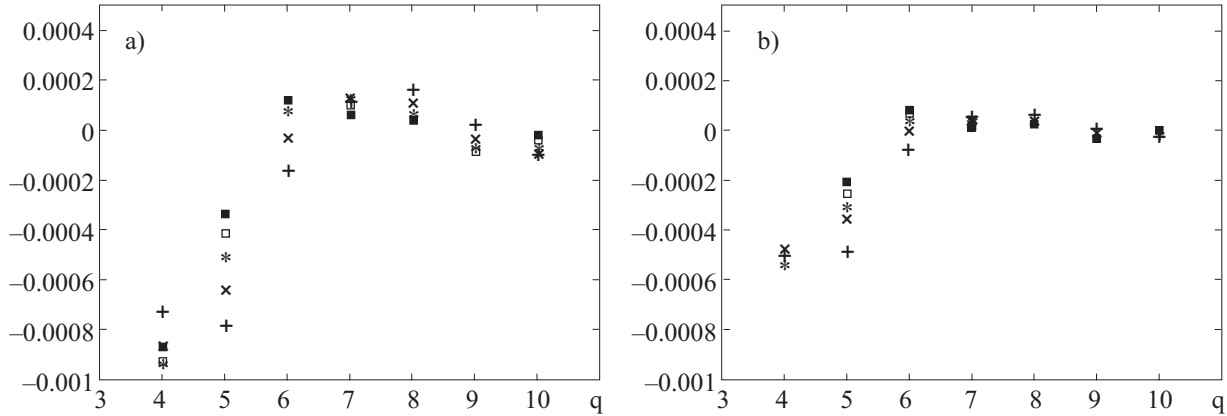, height =5.5cm}}
\caption{\label{fig4} {\small \sl The $H_q$ moments for jets
reconstructed from hadrons for PYTHIA events generated with L3
parameters. In Fig. (a) only charged hadrons are used; in Fig. (b)
all stable particles are counted. Crosses, $x$-s, stars, open and
closed squares correspond to the increasing average number of
particles in a jet (from about 1.1 to 1.5 in Fig. (a) and from 1.3
to 2 in Fig. (b).}}
\end{figure}

  \par
   When we count all the stable particles, we may select the values of
   jet defining parameters
   for which their average multiplicity is the same as the average number of charged stable hadrons. Also
   in this case we observe a striking difference between the values of higher moments for jets
   and hadrons. With decreasing
   average jet multiplicity (and increasing average number of particles in a single jet) even the second
   cumulant (not shown) becomes negative and all the pattern of the dependence of moments $H_q$ on their
   order $q$ does not
   resemble even roughly the pattern found for the single charged stable hadrons.
   \par
   An experimental investigation of the $H_q$ moments for jets has been
   performed by the L3 collaboration. The results are available in
   the D.J. Mangeol Ph.D. Thesis \cite{MA}. Although the jet
   definition used (Durham algorithm) was different than the default
   algorithm from PYTHIA used by us, the results are very similar
   to those presented in Fig. 4a: with increasing "jet size"
   the pattern of moments changes significantly.
   \par
   We regard this dependence as a suggestion that the
   extended local parton-hadron duality (ELPHD), in which one identifies the
   (charged) hadron multiplicity distribution with the parton distribution at small fixed
   virtuality, is not very reliable. We should note,
   however, that the opposite conclusion was drawn from the same
   data \cite{BFO}. The dependence of the $H_q$ moments on their
   order $q$ was investigated there for the parton jets in a MC
   model for varying jet resolution parameter $Q_c$. A  good
   description of L3 data was found for jets as well as for hadrons. This was regarded as a support for ELPHD.
   \par
   The agreement with the L3 data is very good indeed for $Q_c \geq 1GeV$. For small $Q_c$
   the agreement is reasonable, although not perfect. In
   particular, the ELPHD prediction for hadrons is practically
   identical to that for jets with $Q_c=100 MeV$, whereas in data
   there is a marked difference between the moments. This
   difference is similar to that seen for our PYTHIA events: a clear
   minimum for $q=5$ seen for hadrons (Fig. 2) moves already for smallest jets (Fig.4a) so that the values of
   $H_4$ and $H_5$ are almost equal. Thus there seems to be no
   clear "hadron limit" for parton MC, although the differences are not big.

\begin{figure}[h]
\centerline{\epsfig{figure=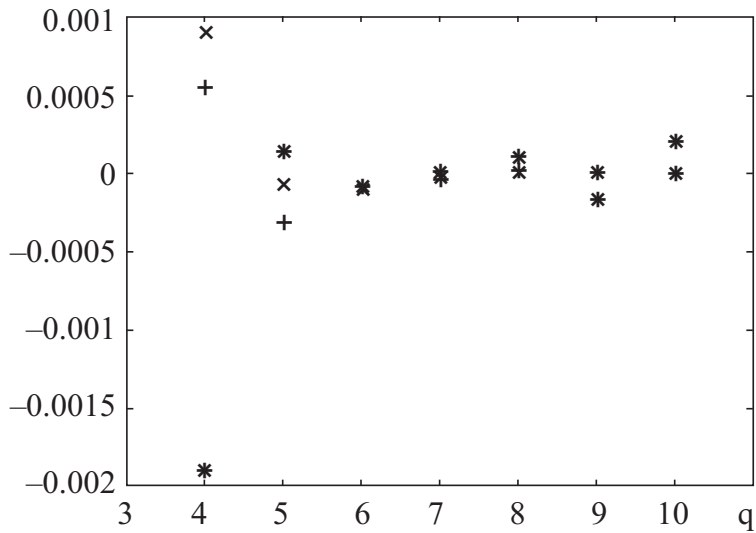, height =7cm}}
\caption{\label{fig6} {\small \sl The $H_q$ moments  for PYTHIA
events generated with L3 parameters. Crosses denote the results
for charged hadrons, $x$-s are for all stable particles and stars
 for positive hadrons.}}
\end{figure}
\par
To verify further the possibility of describing hadron data by
ELPHD we have compared the moments calculated for charged hadrons
(pions, kaons, protons and antiprotons) with the moments
calculated for all stable particles (thus adding photons, neutral
kaons, neutrons and antineutrons). The results, shown in Fig.5,
suggest that a relatively deep minimum at $q=5$ is to a large
extent due to the charge conservation effect. Indeed, the results
for positive hadrons (shown also  in Fig.5) do not exhibit a
pattern seen for all charged hadrons. A deep minimum appears for
$q=4$, and the small oscillations with a very short period follow.
In this case it is rather obvious that the data will confirm our
findings, as the charge conservation is correctly built in PYTHIA
and the charged-neutral correlations are known to be well
described by this generator.
 \par
Obviously, there is no reason why the multiplicity distribution of
partons should be reflected in the distribution of all charged
hadrons more closely than in the distribution of all stable
particles. Thus the observed difference between the moments
calculated for these two choices seems to confirm our doubts
concerning the ELPHD.
\par
The choice of charged hadrons of both signs as a hadronic
counterpart of partons (mostly gluons) in ELPHD is in fact rather
surprising.  The measured cross section for odd multiplicities of
charged hadrons should be zero for an ideal detector, whereas the
distribution of gluons should be smooth without any discrimination
of odd multiplicity values. This suggests strongly that the
agreement of moments calculated for partons and charged hadrons is
to a large extent accidental and is due mostly to the similar
influence of cuts and the two-component mechanism (mentioned in
section 2) in both distributions. In our opinion, one should
rather try to find parameters, for which the parton MC would agree
with the distribution of positive (or negative) hadrons.
\\

\section{Conclusions and outlook}

   Using the PYTHIA
 generator we have investigated
 the multiplicity distributions for hadrons coming from the $Z$ decay. We have calculated the ratios of cumulants
  to factorial moments. The results for charged hadrons in full phase-space are compatible with the L3 data.
    We confirm the earlier findings of a minimum at $q=5,6$ when the moments $H_q$
   are considered as  functions of their order $q$. We confirm also that the truncation of the multiplicity
   distribution causes a significant enhancement of this minimum. However, increasing the statistics of generated
   events beyond $2M$ events does not change visibly the results. This excludes the "natural" truncation due
   to the final statistics as the main source of the minimum.
   \par
   We find that the moments calculated from the multiplicity distributions in the limited part of phase-space
   differ significantly from that for the full phase-space. In particular, the minimum in the dependence of $H_q$
   on $q$ shifts to higher values of $q$ for smaller range of momenta. However, introducing an universal truncation
   on the multplicity distributions for different ranges of momenta, we recover a stable position of the minimum.
   It will be interesting to see if this pattern of generated events will be confirmed in the real data.
   \par
   We have also investigated the moments for the jets, defined as clusters of particles close in momentum space.
   We find that even for very small jets, containing in average only slightly more than one hadron, the pattern of moments
   is different from that for hadrons. The same is true if we use the neutral particles to form jets, for
   which the average multiplicity is the same, as for charged hadrons. Since the distributions of such jets may be
   expected to correspond more closely to the distributions of partons, one may wonder if it is meaningful to
   compare the moments measured for the observed hadrons to the moments calculated for partons. We feel that our
   results cast doubts on the possibility of such an extension of "local parton-hadron duality" beyond the
   average quantities (moments of order one). Again,  more  investigation on the real data would be
   highly desirable.

\section{Acknowledgements}
\par
 We are grateful to A Bia{\l}as and A. Kota{\'n}ski for reading the manuscript and for helpful
 remarks. Fruitful critical remarks by W. Ochs, W. Metzger and by the Referees are gratefully
 acknowledged.

\end{document}